\documentclass[10pt]{article}

\usepackage{latexsym}
\usepackage{amsfonts}
\usepackage{array}
\usepackage[english]{babel}
\usepackage{amsmath}
\usepackage{amssymb}
\usepackage{amsthm}
\usepackage{latexsym}
\usepackage{graphicx}

\def\bs{\boldsymbol}
\def\bE{\begin{equation}}
\def\eE{\end{equation}}
\def\bEA{\begin{eqnarray}}
\def\eEA{\end{eqnarray}}
\def\bEAnn{\begin{eqnarray*}}
\def\eEAnn{\end{eqnarray*}}

\begin{document}
\title{
{\huge \sc Dynamics on/in financial markets:
{\Large \em dynamical decoupling and stylized facts}}\\
\author{S. Reimann$^1$\footnote{Contact: streimann@ethz.ch} \:  \& \: A. Tupak$^2$ \footnote{Contact: tupak@isb.unizh.ch}
\\
$^1$ MTEC - ETH Zurich \\
$^2$ ISB - University of Zurich}}
\maketitle

\begin{abstract}
{\small \em Stylized facts can be regarded as constraints for any modeling attempt of price dynamics on a financial market, in that an empirically reasonable model has to reproduce these stylized facts at least qualitatively. The dynamics of market prices is modeled on a macro-level as the result of the dynamic coupling of two dynamical components. The degree of their dynamical decoupling is shown to have a significant impact on the stochastic properties of return trials such as the return distribution, volatility clustering, and the multifractal behavior of time scales of asset returns. Particularly we observe a cross over in the return distribution from a Gaussian-like to a Levy-like shape when the degree of decoupling increases. In parallel, the larger the degree of decoupling is the more pronounced is volatility clustering. These findings suggest that the considerations of time in an economic system, in general, and the coupling of constituting processes is essential for understanding the behavior of a financial market.}  
\end{abstract}

It is an empirical finding that different financial markets share particular statistical properties. 
Statistical properties that are invariant under the choice of a particular market, are called 'stylized facts', for a survey see \cite{Cont2001} and also the monographs \cite{BouchaudPotters2003, MantegnaStanley2000} as well as the references in them. Stylized facts can be regarded as constraints for any modeling attempt in that a model to be empirically reasonble must exhibit these stochastic properties, at least qualitatively. During the last years a large number of models have been proposed which are able to produce at least some of the stylized facts seen in empirical data. They differ in their respective underlying assumptions. Some of them are based on mainly mathematical assumptions such as stability \cite{Mandelbrot1962} or multifractality \cite{MandelbrotFischerCalvet1997, Bacry2001a}, while others are mostly descriptive in that they consist essentially in postulating that prices follow particular stochastic processes which generate certain distributions such as Levy-processes \cite{MantegnaStanley1994} or hyperbolic processes \cite{Barndorff-Nielsen1978, BibbySorensen1997, Prause1998, Sorensen2003}. These processes often lack a economical basis. Nonetheless they can be extremely valuable concerning forecast, for example. In the following we propose a simple model of price dynamics on a financial market which is oriented at basic economic considerations and analyze this model with respect to the agreement with empirical stylized facts. 

\section{Setting the stage}
The aim of this paper is to model dynamics of prices and to compare simulations results with empiricial data. To avoid misunderstandings already in the very beginning, the model is not about individual traders and their individual decisions. Prices and their dynamics are modelled on the macro level without reffering to any microfoundation. The justification of this approach comes from basic results about huge systems.\\

Prices are endogenously formed in a financial market by the aggregate trading decisions of a large number of investors who interact by trading assets, while the resource 'money' is limited. Prices are therefore regarded as macro-observables of a financial market in a sense which might be comparable to what the pressure in a (classical) gas is or its temperature. In a lower-order approximation the pressure in a gas, for example, is independent from the very nature of the gas molecules including the spatial configuration of the single particles or their mass, for example. Macro-observables of physical systems correspond to so-called {\em Typical Properties} of large systems in the mathematical setting of random systems. These, by definition, are properties almost every micro-realization such a huge system has. The theory of complex systems thus says that micro-details are largely washed out when considering macro-observables. In other words, the theory of huge complex systems suggests that for modelling macrovariables such as prices one can avoid to model details of the micro-level. Particularly our model model of a financial market is not an up-stream analogon of any dynamics on the microlevel concerning concerning individual trading decisions and resulting prices.\\

Although our model is not about individuals, let us personalize the basic model for a moment to give some intuition about what is described. A simple scenario might be the following: There are two financial agents, think about a pension fund and a broker, for example, who both deal with a huge amount of money competing for the same assets. The pension fund is headed by a commitee which meets each quarter to decide about changes in their recent investment strategy according to some news they have received. On the other side there is a broker who makes his decisions about what to buy and what to sell day daily. While in a simple equilibrium model time does not matter, in this model both agents only differ with respect to the time scale on which they alter their respective trading decisions. The pension fund is regarded as 'slow', while the broker is 'fast'.  Both parties compete for assets whose price is determined by trade.  Do these different time scales, their difference respectively, have any impact on the dynamics of prices? This question is answered by simulations. Respective features are not restricted to single parameter values. Simulations are compared to empirical data to demonstrate the significance of these results.\\

The modelling idea is closely related to what is done in the description of multi particle systems: In lower order the overall dynamics of a multi particle is separated into two dynamical components, one is the dynamics of the center of mass, the other one is the dynamics around this center of mass. Note that this corresponds to separating two time scales in the entire system: the dynamics of the center of mass happens slowly, while the dynamics around the center of mass is faster. These two dynamical components are coupled. The gap between the two time scales is known as their dynamic decoupling. Both are coupled giving rise to the overall dynamics of the entire multi particle system.\\

The general idea for modeling a financial market as a multi agent system is the following: We abstract from considering trading individuals and describe the dynamics of prices as the result of the coupling of two {\em dynamical} components. One corresponds to the slowly changing average demand, while the other one takes into account fast fluctuations in demand around this level. Both levels are dynamically decoupled by some degree. The faster fluctuations are the larger is the degree of de-coupling. Then, if prices are determined endogenously related to their excess demand, price dynamics also is described by fluctuations around some level which results from a 'constant' mean demand in assets. Particularly we neglect any drift component in this model. Formally, we decompose the price ${\bs S}_t$ into two parts, a constant component $\overline{\bs S}$ and a fluctuating one ${\bs \sigma}_t$
\bE\label{S1}
{\bs S}_t = \overline{\bs S} \; + \; {\bs \sigma}_t.
\eE
A question, we have to answer, is about the nature of the term ${\bs \sigma}_t$ thus about about corresponding statistical properties of price changes, such as log returns
\bE
{\bf \rho}_{t+1} = \ln \bigl( {\bf S}_{t+1} - {\bf S}_{t} \big) = \ln \frac{{\bs \sigma}_{t+1}}{{\bs \sigma}_{t}}
\eE\\

The more general question is the following: {\em How will the degree of dynamic de-coupling affect the fluctuation of prices of the assets? What are the impacts on th statistical properties of returns?}\\

It will turn out that the dynamical decoupling, i.e. the difference between the two dynamical levels 'slow' and 'fast', has an tremendous impact on the statistical properties of returns including the fatness of tails of the return distribution and the degree of so-called volatility clustering. This finding holds for a wide parameter range and thus is not restricted to only particular parameter setting. In other words, it is a generic property. While time does not play any role in the General Equilibrium Theory of economic systems, time plays an important role in any non-equilibrium situation. 

%Hence dynamtime scales and, more generally, the importance of temporal considerations, not existent in most standard economic theorie, turns out of be of great importance. 

\section{The general model of price dynamics}

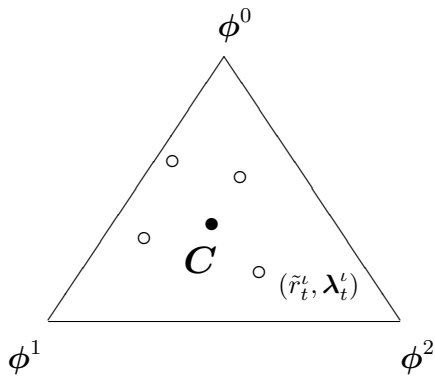
\begin{figure}[h]
\setlength{\unitlength}{0.9cm}
\begin{center}
\begin{picture}(4.0,5)
\thinlines
%\put(0.0,0.0){\dashbox{0.2}(4.5,4.5)}
\put(-0.3,0.08){\line(1,0){5.2}}
\put(2.3,4){\line(2,-3){2.6}}
\put(2.3,4){\line(-2,-3){2.6}}
\put(2.2,4.3){\large $\bs{\phi}^0$}
\put(4.9,-0.6){\large $\bs{\phi}^2$}
\put(-0.9,-0.6){\large $\bs{\phi}^{1}$}
\put(2.42,2.1){\large $\circ$}
\put(3.1,0.5){($\tilde{r}^\iota_t,\bs{\lambda}^\iota_t)$}
\put(1.42,2.34){\large $\circ$}
\put(1.7,0.8){\Large $\bs{C}$}
\put(2.7,0.7){\large $\circ$}
\put(2,1.4){\large $\bullet$}
\put(1,1.2){\large $\circ$}
\end{picture}
\end{center}
{\caption{\label{simplex} {\footnotesize \em A financial market as a multi particle system: The population of agents $\{\iota\}$ in a financial market is displayed a cloud of points $\bs{\lambda}^\iota_t$ with masses $\tilde{r}^\iota_t$ in the simplex spanned by the fundamental investment styles $\{\bs{\phi}^0,\bs{\phi}^1,\bs{\phi}^2\}$.}}}
\end{figure}

The model is not about individual investor rather than about investment strategies. This picture traces back to the formal setting outlined in \cite{BlumeEasley1992}. An investment strategy is a rule which determines how an investor distributes his wealth over available assets. This also includes a riskless asset or a bank account. Investors who invest all their money in the financial market are fully invested. Instead of considering the set of individual investors, we consider the set of investment strategies on the market, while the weight of a strategy is the amount of money invested on the financial market according to this strategy. Since investment strategies interact by competing for the same risky assets, the financial market in this picture is an interacting particle system in which 'particles' are now investment strategies rather than individuals or institutions, see Figure \ref{simplex}.\\ 

%
%This particularly means that the investor decides about the amount of money he is willing to carry to the financial market in the very beginning and let this money work without ever altering the amount of invested money 

Given that there are $n+1$ assets $k = 0, .., n$ available on the market the agents are competing for. The riskless asset $k=0$ can be regarded as a bank account on which investors can park cash.
In the setting supposed financial agents are investment strategies ${\bs \lambda}^\iota = \bigl( \lambda^\iota_k\bigr)_{k=0}^n$, where $\lambda^\iota_k$ is the portion of recent wealth $m^\iota$ agents $\iota$ invests into assets $k$ at that time. Thus if $\lambda^\iota_0 > 0$, the investor is not fully invested on the financial market.
The individual demand in asset $k$ then is $m^\iota \: \lambda^\iota_k$, while the aggregate demand is the sum over individual demands. It follows that
\bE\label{L}
1 \; = \; \sum_{k\ge 0} \lambda^\iota_k \; = \; \lambda^\iota_0 + \sum_{k > 0} \lambda^\iota_k
\eE

Each risky asset has a price endogenously determined in the market according to the excess demand in this asset. The following observation is concerned with the (Walrasian - like) 'mechanism' how prices are set on a market. The market considered is 'closed' in that, at any time, the total amount of assets is preserved, i.e. during trade only the distribution of assets over the investors' population is changed, while neither new assets enter the market nor old ones are removed from it. 
\\
%\newpage

{\bf Observation:} {\em If the number of units of all risky assets is conserved during trade, their prices yield 
\bE\label{MCP}
\tilde{\bs S}_t \; =\; \sum_\iota \tilde{r}^\iota_t \; {\bs \lambda}^\iota_t .
\eE
where $\tilde{r}^\iota_t$ is the relative wealth of agent $\iota$ on the financial market. Since $\| \tilde{\bs S}_t\| = 1$ these prices are called {\sc relative prices.}}\\

\proof{ 
Given the recent price system ${\bs s}_t$ on the financial market, agent $\iota$ builds his portfolio\footnote{We use the following notation: For two vector ${\bs x} = (x_k)$ and ${\bs y} = (y_k)$, ${k=1..K}$, $\frac{\bs x}{\bs z} = \left( \frac{x_k}{y_k}\right)$ denotes component-wise division, while ${\bs x} \star {\bs y} = \bigg( x_k \: y_k\bigg) $ denotes component-wise multiplication. Finally ${\bs x} \bullet {\bs y} \equiv {\bs x}{\bs y} = \sum_k x_k y_k$ is the standard product.} 
$
{\bs \theta}^ \iota_t \; = \; \frac{m^ \iota_t \; {\bs \lambda}^ \iota_t}{{\bs s}_t},
$ 
where $\theta^ \iota_{t,k}$ is the number of units of assets $k$, agent $\iota $ buys for his wealth at time $t$ for its recent price $s_{t,k}$. Then the total number of units of a risky asset $k>0$ is $\sum_ \iota \theta^ \iota_{t,k}$. Conservation of assets then means that
$
\sum_ \iota {\bs \theta}^ \iota_t = \sum_\iota {\bs \theta}^ \iota_{t+1}
$
for all $t$. Denoting the aggregate demand by ${\bs \delta}_t = \sum_ \iota m^\iota_t {\bs \lambda}^ \iota_t$ we thus obtain 
$
\frac{ {\bs \delta}_t}{{\bs s}_t} = 
\frac{{\bs \delta}_{t+1}}{ {\bs s}_{t+1}}
$
which is solved by 
$
{\bs s}_t =  {\bs \delta}_t \star \tilde{\bs s}_0$, where $\tilde{\bs s}_0 = \frac{{\bs s}_0}{{\bs \delta}_0}$. Without loss of generality we put the net supply $\tilde{\bs s}_0 = {\bs 1}$, hence we arrive at
$
{\bs s}(t) = \sum_ \iota m^ \iota_t {\bs \lambda}^ \iota_t
$, i.e. prices are due to the aggregate demand.
Then, according to eq. \ref{L}, the aggregate demand in risky assets yields $\|{\bs s}_t\| = \sum^\iota m^\iota_t (1 - \lambda^\iota_0) = m_t - \hat{m}_t$, where $m_t = \sum_\iota m^\iota_t$ is the total wealth, while $\hat{m}_t = \sum_\iota m^\iota_t \lambda^\iota_{0,t}$ is the wealth put in the bank account. Defining $\mu_t := \frac{\hat{m}_t}{m_t}$, we obtain $\|{\bs s}_t\| = m_t (1-\mu_t)$. Finally, by defining the relative wealth of agent $\iota$ on the financial market by $r^\iota_t = \frac{1}{(1+\mu_t)} \; \frac{m^\iota_t}{m_t}$, we derive the result from ${\bs S}_t = \frac{{\bs s}_t}{\|{\bs s}_t\|}=  \sum_\iota \tilde{r}^\iota_t {\bs \lambda}^\iota_t$. Finally it is easily checked that $\|\tilde{\bf S}_t\| = \sum_{k>0} \tilde{S}_{t,k} = 1$.}\\

While it is realistic to say that most investors hold some position in cash, it may also be realistic to assume that the cash position of investors on the financial market is still small compared to the sum of risky positions. 
Note that if all investor is fully invested, i.e. $\lambda^\iota_{0,t} = 0$ for all $\iota$, then $\mu_t = 0$ and $r^\iota_t = \frac{m^\iota}{m_t}$. We define corresponding relative prices by
${\bf S}_t \; = \; \sum_\iota r^ \iota_t \; {\bs \lambda}^ \iota_t$. Note that for  a small degree of cash investment, i.e. $\mu_t \ll 1$ we obtain
\bE
\tilde{\bs S}_t = {\bs S}_t \; \bigl( 1 + \mu_t + {\cal O}(\mu_t)\Bigr)
\eE
In this sense ${\bf S}_t$, i.e. prices determined under the condition that all investor are fully invested, can be used as a first order approximation. Moreover, results will be largely uneffected except that the perturbation term $\mu_t$ introduces a second source of randomness\\

%\newpage

The following remarks might be in place. Concerning economic thinking it may help intuition to recognize that these prices are identical with so-called 'market-clearing prices'. From a more physical point of view it might be helpful to observe the following: A financial market can be viewed as a collection of point-masses $\iota$ with coordinates ${\bs \lambda}^\iota_t$ having weight $r^\iota_t$ at time $t$. Then the center of mass yields ${\bs C}_t = \frac{\sum_\iota m^\iota_t \: {\bs \lambda}^\iota_t}{\sum_\iota m^\iota_t} = {\bs S}_t$ for all $t$. This justifies to regard prices as average properties of the system.

We now assume that the price process ${\bs S}_t$ can be described by the interaction of two components, in the following called 'agents' for short. To avoid misunderstandings, in our model {\em agents are not individuals}. In a low order approximation we describe price dynamics generated by the interplay of only two dynamical components, while for higher order approximations more components should be taken into account.
Agents $'a,b'$ have wealth $r^a_t$ and $r^b_t$, respectively, and follow the investment strategies ${\bs \lambda}^a$ and ${\bs \lambda}^b_t$. Since $r^a_t + r^b_t = 1$, we arrive at
\bE\label{Sr}
{\bs S}_t \; = \; {\bs \lambda}^a \; + \; \big( {\bs \lambda}^b_t - {\bs \lambda}^a \big) \: r^b_t.
\eE
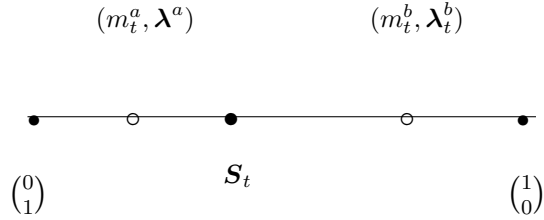
\begin{figure}[h]
\setlength{\unitlength}{1.3cm}
\begin{center}
\begin{picture}(6.0,3)
\thinlines
%\put(0,0.0){\dashbox{0.2}(5,2)}
\put(0,1.08){\line(1,0){5.2}}
\put(4.9,0.2){\large ${1 \choose 0}$}
\put(-0.2,0.2){\large ${0 \choose 1}$}
\put(0.7,2){$(m^a_t, \bs{\lambda}^a)$}
\put(3.5,2){$(m^b_t,\bs{\lambda}^b_t)$}
\put(2,0.4){$\bs{S}_t$}
\put(3.8,0.98){\large $\circ$}
\put(2,0.98){\large $\bullet$}
\put(5,0.98){$\bullet$}
\put(0,0.98){$\bullet$}
\put(1,0.98){\large $\circ$}
\end{picture}\par
\parbox{10.5cm}{
\caption{\label{simplex2} {\footnotesize The population of agents $\{a,b\}$ in a financial markets with two assets is a pair of points $\bs{\lambda}^\iota_t$ with 'masses' $m^\iota_t$, $\iota=a,b$ in the 1-simplex spanned by the fundamental investment styles ${1 \choose 0}, {0 \choose 1}$. The corresponding 'market-clearing' price is ${\bs S}_t$ and corresponds to the center of mass of the system.}}
}
\end{center}
\end{figure}

\subsection{The wealth process}
We now have to model the wealth process of agent $'b'$. This is done similar to the Lukas model for so-called short lived assets. According to this model, the agent $\iota$ is characterized by the pair $(m^\iota_t, {\bs \lambda}^\iota_t)$ and buys his portfolio on the financial market ${\bs \theta}^\iota_t = m^\iota_t \: \frac{{\bs \lambda}^\iota_t}{{\bs s}_t}$ for some price ${\bs s}_t$. If one unit of asset $k$ has some future value $D_{k,t+1}$ tomorrow, then the future value of the portfolio is $ {\bs \theta}_t \: {\bs D}_{t+1}$. His wealth therefore evolves according to 
$m^\iota_{t+1} = m^\iota_t \:  \left( \frac{ {\bs D}_{t+1}}{{\bs s}_t} {\bs \lambda^\iota_t}\right)
$
%\bEAnn
%m^\iota_{t+1} &=& m^\iota_t \:  \left( \frac{ {\bs D}_{t+1}}{{\bs s}_t} {\bs \lambda^\iota_t}\right)
%\\
%\frac{m^\iota_{t+1}}{m_{t+1}} &=& \frac{m^\iota_t}{m_t} \:  \left( \frac{m_t}{m_{t+1}}\frac{ {\bs D}_{t_+1}}{{\bs s}_t} {\bs \lambda^\iota_t}\right)\\
%r^\iota_{t+1} &=&  \frac{1}{m_{t+1}}\frac{ {\bs D}_{t+1}}{{\bs S}_t} \left( r^\iota_t \:{\bs \lambda^\iota_t}\right)\\
%1 &=& \frac{1}{m_{t+1}} \sum_k D^k_{t+1}\\
%m_{t+1} &=& \|{\bs D}_{t+1}\|_1
%\eEAnn
Note that $\|{\bs D}_{t+1}\|_1 = m_{t+1}$. Then, 
transforming to relative variables $r^b_t = \frac{1}{1+\mu_t}\frac{m^b_t}{m_t}$, equivalently gives
\bE
r^b_{t+1} \; = \; r^b_t \: \left( \frac{{\bs d}_{t+1}}{{\bs S}_t} {\bs \lambda}^b_t\right),
\eE
where ${\bs d}_{t+1} = \frac{{\bs D}_{t+1}}{\|{\bs D}_{t+1}\|_1}(1+\mu_t)$, and $\tilde{\bs S}_t$ are the relative prices. The relative wealth of agent $'b'$ therefore has an (uncertain) growth rate 
\bEA\label{r}
\beta_{t} &=& \frac{r^b_{t}}{r^b_{t-1}}
%&=& \left( {\bs d}_{t} \; \frac{{\bs \lambda}^\iota_{t-1}}{{\bs S}_{t-1}}\right) \: r^\iota_{t-1}\\
= \frac{1}{1+\mu_t}\frac{{\bs d}_{t}}{{\bs S}_{t-1}} \; {\bs \lambda}^b_{t-1} .
\eEA
Note that the growth rate particularly depends on the price system of the market in that it is linear in $ \frac{\bs 1}{{\bs S}_{t-1}}$.
This establishes a multiplicative stochastic negative feedback in price dynamics. If the agents are not fully invested the growth rate receive a second source of randomness which is due to the fluctation of money invested on the financial market. Corresponding to eq. \ref{MP} this pops up in an additional multiplicative term in the growth rate. As seen below this particular structure generates important features of statistical properties in this market. \\

The price process consequently becomes
\bE
{\bs S}_{t} \;=\; {\bs \lambda}^a \; + \; ({\bs \lambda}^b_{t} - {\bs \lambda}^a) \; \prod_{\tau=-1}^t \;\beta_\tau, \qquad \beta_{-1} = r^b_0.
\eE
This equation has the form ${\bs S}_t = \overline{{\bs S}} + {\bs \sigma}_t$. While the price process has a constant component, $\overline{\bs S} = {\bs \lambda}^a$, prices fluctuate around this level according to the fluctuating demand generated by agent $'b'$, given by ${\bs \sigma}_t = ({\bs \lambda}^b_{t} - {\bs \lambda}^a) \; \prod_{\tau=-1}^t \;\beta_\tau$. Roughly speaking one can say that price dynamics is essentially driven by the expected growth rate of agent $'b'$. Interaction of agents $'a,b'$ comes from two sources: one is ${\bs \lambda}^b_t - {\bs \lambda}^a$, while the other one is due to the pricing formula ${\bs S}_t =r^a_t {\bs \lambda}^a + r^b_t {\bs \lambda}^b_t$, see equation \ref{MCP}.

\subsection{An asymptotic result}

Price dynamics follows the wealth accumulation of agents $'b'$ as well as from the interplay of the two strategies ${\bs \lambda}^a$ and ${\bs \lambda}^b_t$. The following result singles out a special strategy which is growth optimal in that it collects the entire wealth of the market asymptotically \cite{Amir2005}. This result is based on the assumption that the value process is stationary, i.e.
$$
{\bs d}_t \sim {\cal F}_{\bs d}
$$
whose first moment exists. In this case relative prices become asymptotically constant, while convergence is exponential. In the following we give a simple heuristic argument for this result.\\

{\bf Observation:} {\em 
If ${\bs d}_t \sim {\cal F}$ and ${\bs \lambda}^a = {\mathbb E}_{\cal F}[{\bs d}]$, then ${\bs S}_t \to {\mathbb E}_{\cal F}[{\bs d}]$ for $t \to \infty$ exponentially.
}\\

\proof{\em
Assume that agent $'a'$ has nearly overtaken the market, i.e. $r^a_t = 1 - \epsilon_t$. Then ${\bs S}_t = {\bs \lambda}^a + {\cal \bf O}(\epsilon_t)$. Therefore the growth rate yields
$
\beta_t = \frac{{\bs d}_t}{{\bs \lambda}^a} {\bs \lambda}^b_t,
$.
where now $\beta_\tau$ are iiid random variables. It follows from $\beta_t = \prod_{\tau=0}^{t-1} \beta_\tau$ that $r^b_t \sim r^b_0 e^{t\:
{\cal H}_{{{\cal F}_d}}({\bs \lambda}^b)}$, where the entropy growth rate obeys
%\bEAnn
$
{\cal H}_{{\cal F}_{\bs d}}({\bs \lambda}^b)  =  {\mathbb E}_{{{\cal F}_{\bs d}}} \: \ln \:\left[ \frac{{\bs d}_t \star {\bs \lambda}^b_t}{{\bs \lambda}^a}\right] \;
%< \; \ln \; {\mathbb E}_{{\cal F}_{\bs d}} \left[ \frac{{\bs d}_t \star {\bs \lambda}^b_t}{{\bs \lambda}^a}\right] \;
%\le 
< \ln \: {\mathbb E}_{{\cal F}_{\bs d}} \left[ \frac{{\bs d}_t }{{\bs \lambda}^a}\right]
%\eEAnn
$. 
Therefore if ${\bs \lambda}^a = {\mathbb E}_{{\cal F}_{\bs d}}[{\bs d}]$, then ${\cal H}_{{\cal F}_{\bs d}}({\bs \lambda}^b) < 0 $ and $r^b_t \to 0$ exponentially. Then
$
{\bs S}_t \to {\bs S} = {\mathbb E}_{{\cal F}_{\bs d}}[{\bs d}]
$
exponentially. 
}\\

\begin{figure}[h] 
\setlength{\unitlength}{0.8cm}
%\hskip -0.5cm
\begin{minipage}[t]{8 cm}
	\includegraphics[width=73mm]{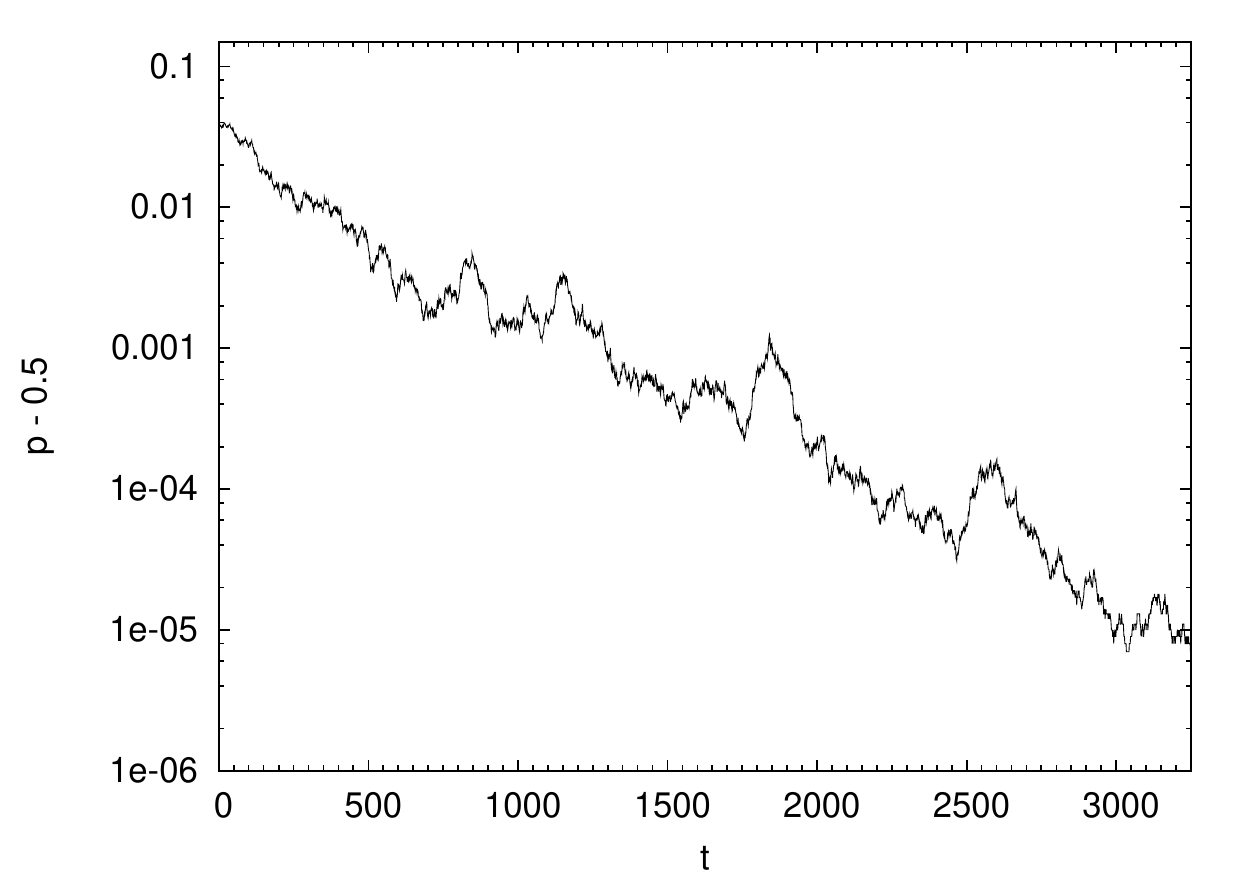}\par
	\parbox{70mm}{ \caption{\label{sstrail}\footnotesize Exponential convergence of the relative price ${\bs S}_1 = p$ to $1/2$ in a market with two constant strategies $\bigg( [1/2][0.40]\bigg)$, see below} }
 \end{minipage} %\hskip 0.3cm
\begin{minipage}[t]{8 cm}
	 \includegraphics[width= 73mm]{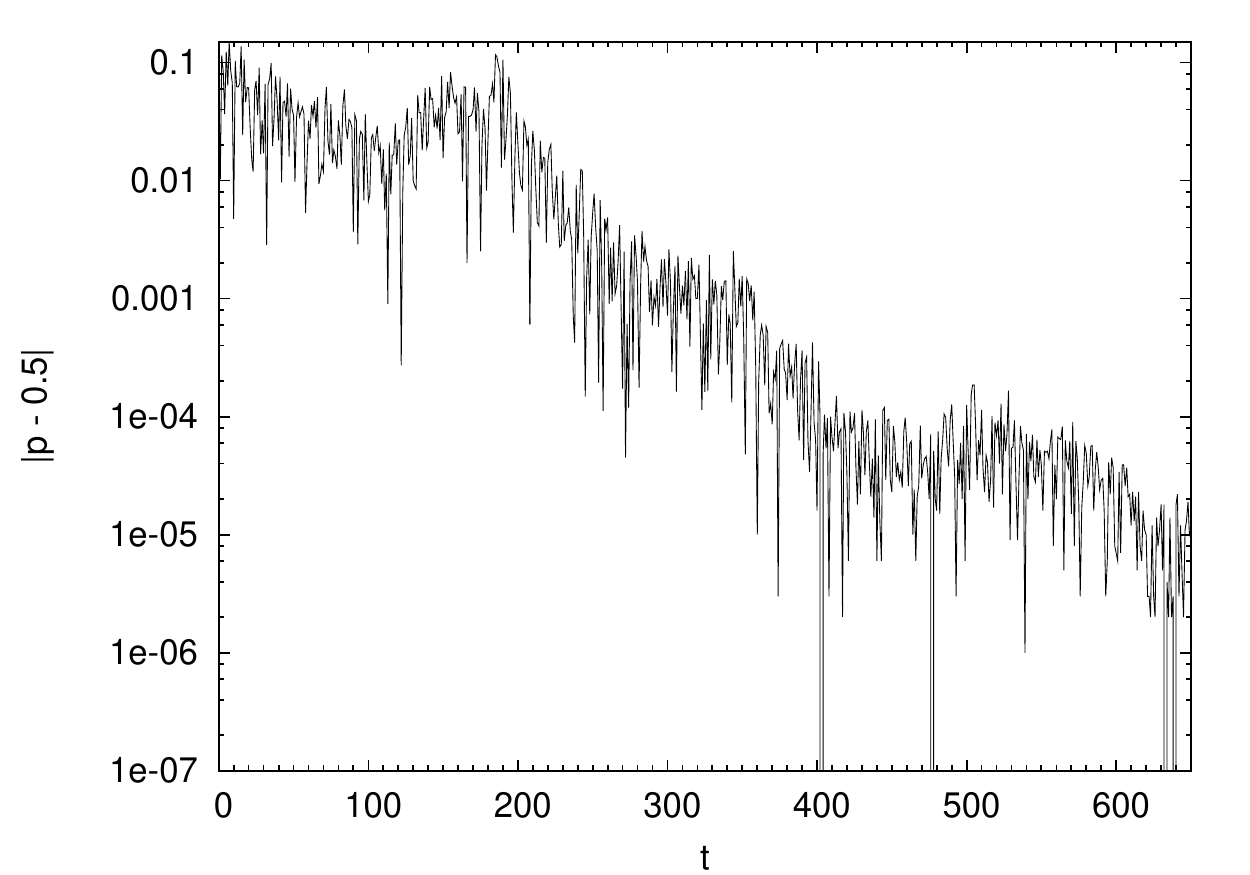}\par
	\parbox{70mm}{ \caption{\label{snstrail}\footnotesize Convergence of relative prices ${\bs S}_1 = p$ in a market with two strategies$\bigg( [1/2], (0.45)\bigg)$, i.e. a onstant one and a fluctuating one.} }
 \end{minipage} 
 %\end{center}
\end{figure}

The following pictures \ref{sstrail} and \ref{snstrail} show the exponential convergence of prices in a semi-logarithmic plot. For notation see eq. \ref{notation}. Due to the exponential convergence of prices, the distribution of log returns will converge to a Dirac function with mass in $0$ rapidly. 

\section{Setting the stage for the simulations}

In the following we display simulations of the price dynamics generated by the interaction of two agents $'a,b'$ competing for two assets and analyze statistical properties of corresponding return trails. Recall that agent $'a'$ represents the average component of the dynamics with relative wealth $r^a_t$ and a constant investment strategy ${\bs \lambda}^a$, while the other one, $'b'$ has wealth $r^b_t$ and follows a dynamical investment strategy ${\bs \lambda}^b_t$. Since ${\bs \lambda}^b_t$ shall represent fluctuations in demand around some level, it is reasonable to make its dynamics random itself. The proposed setting is as follows:
\bEA
{\bs \lambda}^a &=& {a \choose 1- a}, \qquad 0 \le a \le 1,\\
{\bs \lambda}^b_t &=& \frac{\bs 1}{\bs 2} + \frac{b}{2} {1 \choose -1}\: x_t, \qquad
0 \le b \le 1 
\eEA 
where $x_t$ is uniformly distributed in $[-1,+1]$. In other words, in the average agent $'b'$ regards both assets are identical. The larger the parameter $b$ is, the stronger are the fluctuations caused by this component.
 For notational convenience we write
 \bE\label{notation}
 [a] \; \equiv \;  {\bs \lambda}^a , \qquad (b) \; \equiv \;  {\bs \lambda}^b_t. 
 \eE
In a recent paper, the situation was considered in which the flexible strategy $(b)$ receives information about the excess return of the two assets and thus reacts to their performances \cite{Reimann2007a}.\\
 
Finally we have to fix the properties of the value process ${\bs d}_t$. In this model wealth dynamics is exclusively driven by the value process ${\bs d}$ of assets, see eqn \ref{Sr}. The future value of an asset is uncertain and subject to large number of factors. Undoubtedly these factors and their interplay as well as the structure of the financial market it self vary over time. It is a common assumption that whatever evolution takes place, this happens sufficiently slowly compared to price dynamics. Then, as an approximation, the hypothesis may be justified that the distribution of this process is stationary. We therefore assume that ${\bs d}_t \sim {\cal F}_{\bs d}$. Adopting a Baysian view point ${\cal F}_{\bs d}$ represents the knowledge we have about the value process. From the construction we know that ${\bs d}_t = {\delta_t \choose 1 - \delta_t}$, where $0 \le \delta_t \le 1$. In order not to implement more knowledge about this random variable, i.e. to stay with a minimal of pre assumptions we choose the distribution having minimal information on a finite interval, which is the uniform one
\bE\label{F}
\delta_t \sim {\cal U}[0,1].
\eE
This choice reflects minimal prejudice rather than economic knowledge. Model improvements would also consist in drawing more information about the respective process and to postulate an other distribution. In the following we study price dynamics in this basis setting.

%\begin{figure}[h] 
%\setlength{\unitlength}{0.8cm}
%\begin{center}
%\begin{minipage}[t]{9 cm}
%	 \includegraphics[width= 90mm]{test_s50_c30_histo}\par
%	\parbox{90mm}{ \caption{\label{s50c30Hist}\footnotesize Distribution of log-returns on a market $([1/2], 0.45)$ } }
% \end{minipage} 
%\end{center}
%\end{figure}

Note that in this case ${\mathbb E}_{{\cal F}_{\bs d}}[{\bs d}] = \frac{{\bs 1}}{{\bs 2}}=[1/2] = (0) $. Thus, if the parameters $a=1/2$ or $b=0$, prices converge exponentially towards $\frac{{\bs 1}}{{\bs 2}}$, while the distribution of price changes converges to a Dirac function rapidly. This situation is shown in Figure \ref{s50c30Hist}. Comparing related return trails with empirical ones makes it obvious that our model with these parameter settings does not correspond to empirical financial markets' data. For the simulations we thus exclude these parameter settings from our simulations.\\

Strategy parameters $a$ and $b$ can take values in $(0,1)$ and $[-1,+1]$ respectively. On the other hand our model contains some degree of symmetry in that both assets are treated as being essentially the same. This allows to restrict ourselves to the parameter ranges
\bE
0 < a < \frac{1}{2} \qquad \wedge \qquad 0 < b \le 1
\eE
in the simulations. 

\section{The simulations}

We simulate price dynamics generated by the interplay of the two strategies $\bigg( [a], (b) \bigg)$. Since $\|{\bs S}_t\|_1 = S^1_t + S^2_t = 1$, relative prices have the same statistical properties. Let $\tilde{Z}_t = \ln \frac{S^1_{t+1}}{S^1_{t}}$ be the log-returns, while 
$$
Z_t \; = \; \frac{\tilde{Z_t} - \langle Z \rangle}{\sigma(Z)}
$$
are the standardized returns. We estimate trails of length $5.000$, which corresponds to daily data over a time span of approximately 20 years. Statistical properties considered are 
\begin{itemize}
\item the autocorrelation of integer powers of absolute returns, see subsection \ref{autocorr}
\bE
C_\alpha(\tau) = corr\bigg( |Z_{t+\tau}|^\alpha, |Z_t|^\alpha\bigg), \qquad \alpha = 1,2, ...
\eE
\item the multiscaling spectrum indicating multifractal behavior in time series of return trails, see subsection \ref{multifract} , and finally
\item the relative frequencies of log returns, see subsection \ref{distribution}
\bE
f_Z(z) = {\mathbb P}[Z_t = z]
\eE
\end{itemize}
How do these quantities vary when the strength of the internal fluctuation, i.e. $|b|$, increases from $0$? \\

{\bf Main observations are}: {\em The stronger internal fluctuations are, i.e. the larger $|b|$ is, the more pronounced is 'volatility clustering' in log-returns, i.e. the slower is the decay of the autocorrelation of absolute returns; The multifractal spectrum becomes more non-linear; Increasing $|b|$ leads to a cross over from a concave to a convex shaped distribution.}\\

\begin{figure}[h] 
\setlength{\unitlength}{0.8cm}
%\begin{center}
\begin{minipage}[t]{7 cm}
	\includegraphics[width=70mm]{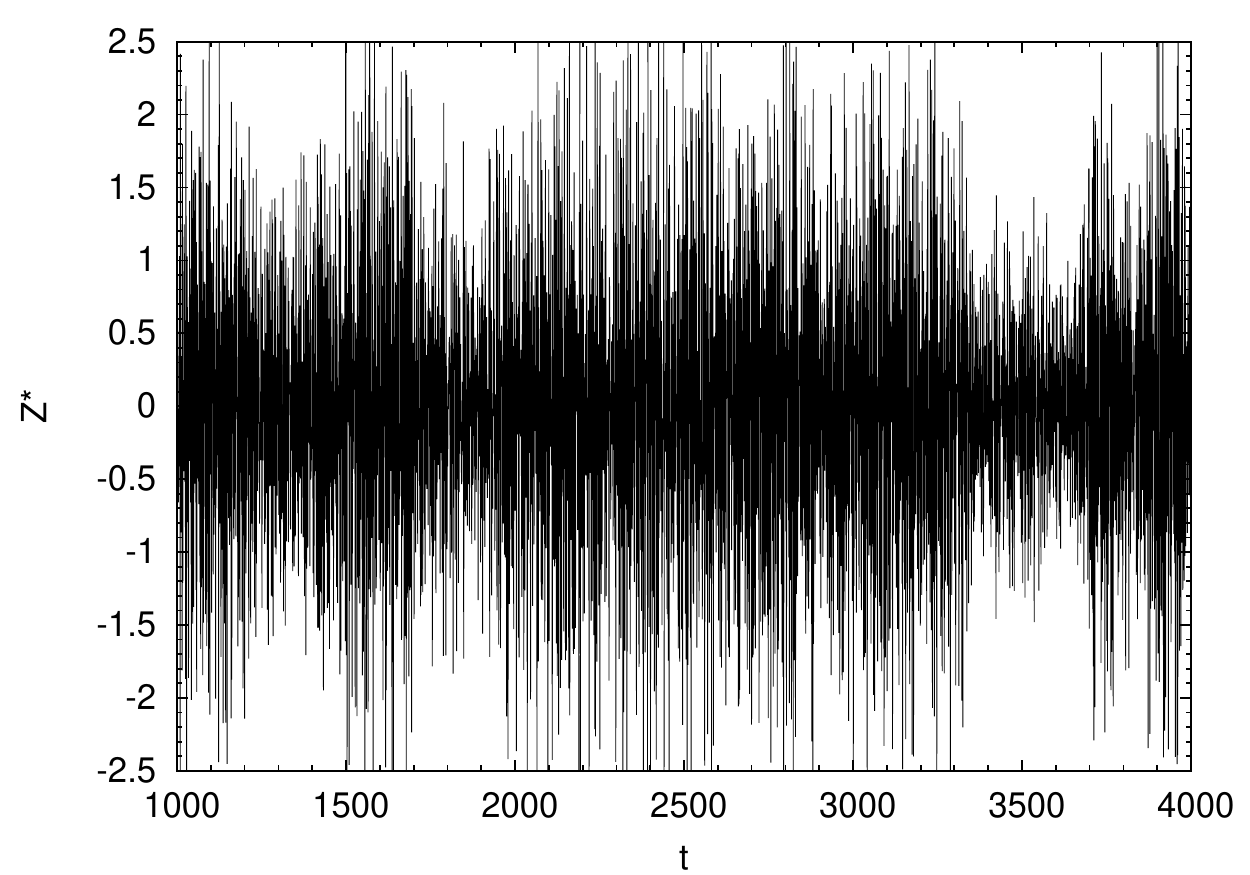}\par
	\parbox{50mm}{ \caption{\label{4025trail}\footnotesize Return trail of the market $([0.40], (0.25))$} }
 \end{minipage} \hskip 0.1cm
\begin{minipage}[t]{5 cm}
	 \includegraphics[width= 70mm]{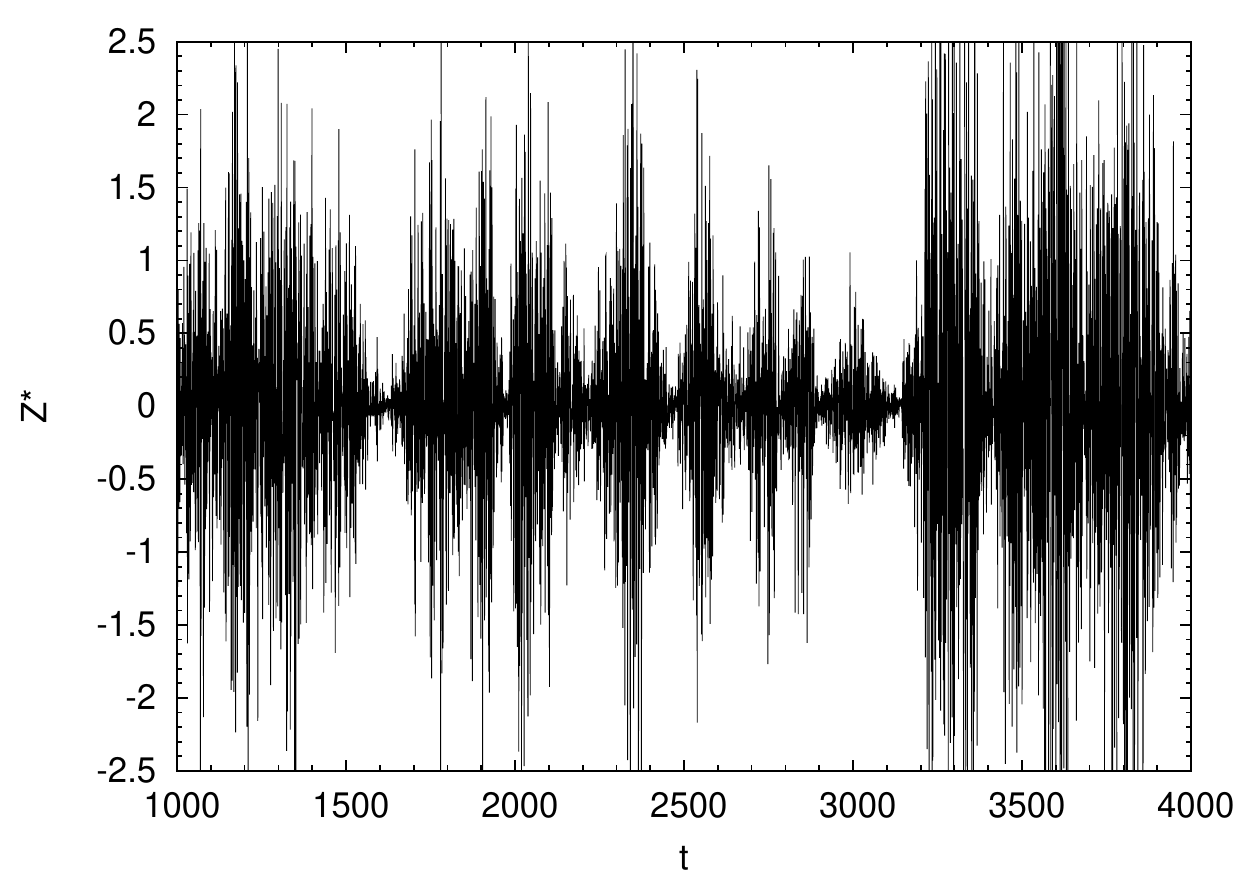}\par
	\parbox{50mm}{ \caption{\label{4045trail}\footnotesize Return trail of the market $([0.40], (0.45))$} }
 \end{minipage} 
 \par  
 \begin{minipage}[t]{7 cm}
	\includegraphics[width= 70mm]{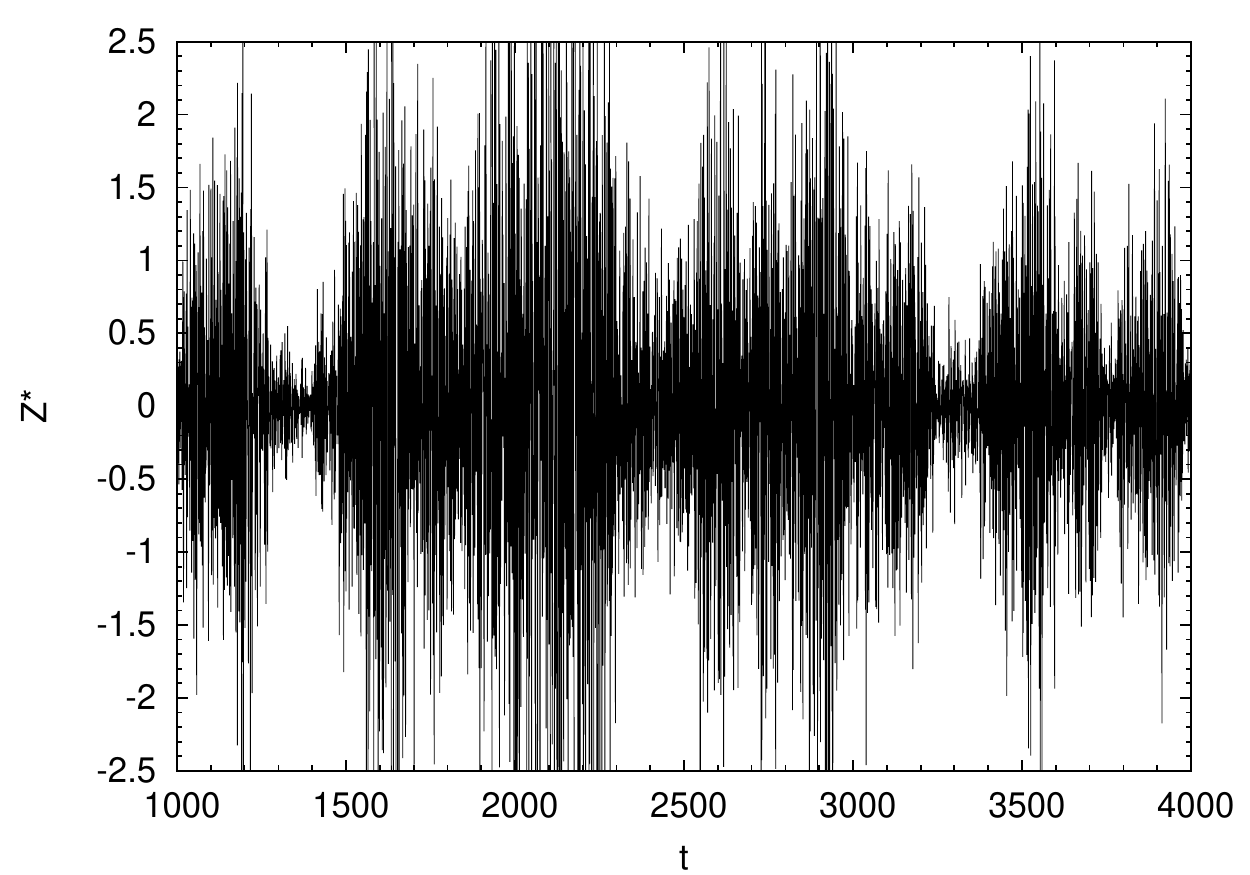}\par
	\parbox{50mm}{ \caption{\label{4525trail}\footnotesize Return trail of the market $([0.45], (0.25))$} }
 \end{minipage} \hskip 0.1cm
\begin{minipage}[t]{7 cm}
	 \includegraphics[width= 70mm]{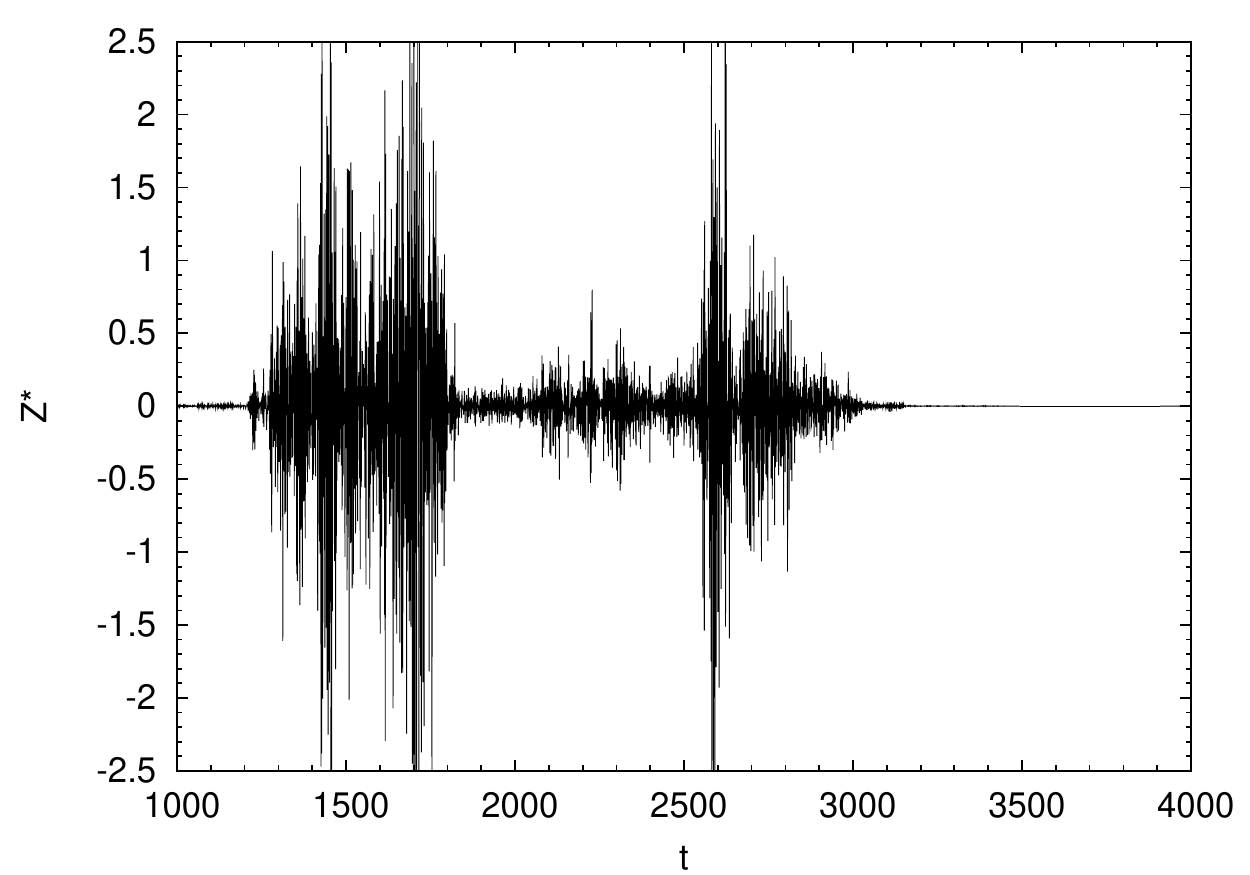}\par
	\parbox{50mm}{ \caption{\label{4545trail}\footnotesize Return trail of the market $([0.45], (0.45))$ } }
 \end{minipage} 
 %\end{center}
\end{figure}

We start by showing some typical return trails for fixed $[a]=[0.40]$, while $(b)$ runs over the considered range as indicated, see Figures \ref{4025trail} to \ref{4545trail}. As as first rough observation: The large the parameter $|b|$ is, i.e. the stronger the system internal fluctuations are, the more pronounced is 'volatility clustering'. \\

\subsection{'Volatility clustering': Decay of auto-correlations}\label{autocorr}

\begin{figure}[h] 
\setlength{\unitlength}{1cm}
\begin{center}
\begin{minipage}[t]{7 cm}
	\includegraphics[width= 70mm]{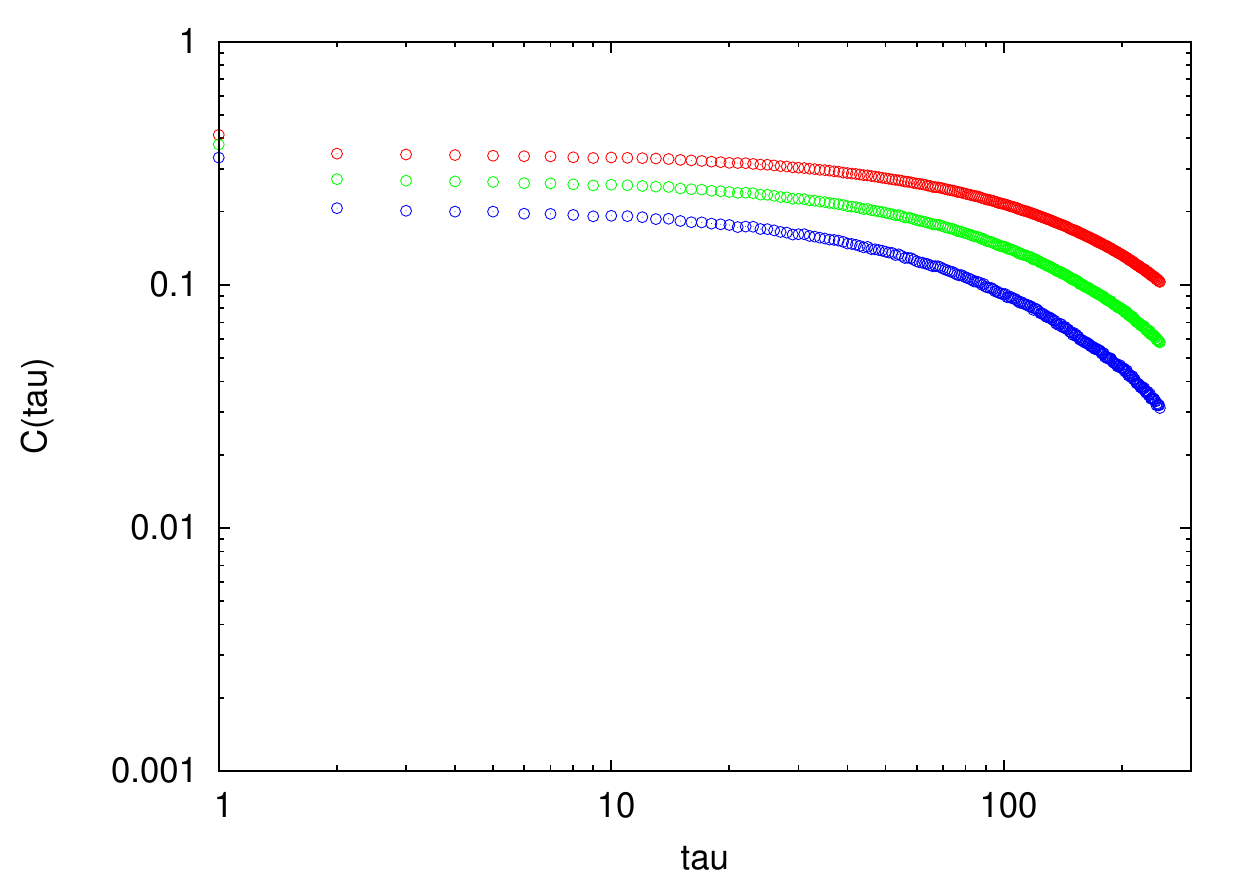}\par
	\parbox{70mm}{ \caption{\label{4525acor}\footnotesize $\langle C_\alpha(\tau) \rangle$ for $([0.45], (0.25))$} }
 \end{minipage} \hskip 0.1cm
\begin{minipage}[t]{7 cm}
	 \includegraphics[width= 70mm]{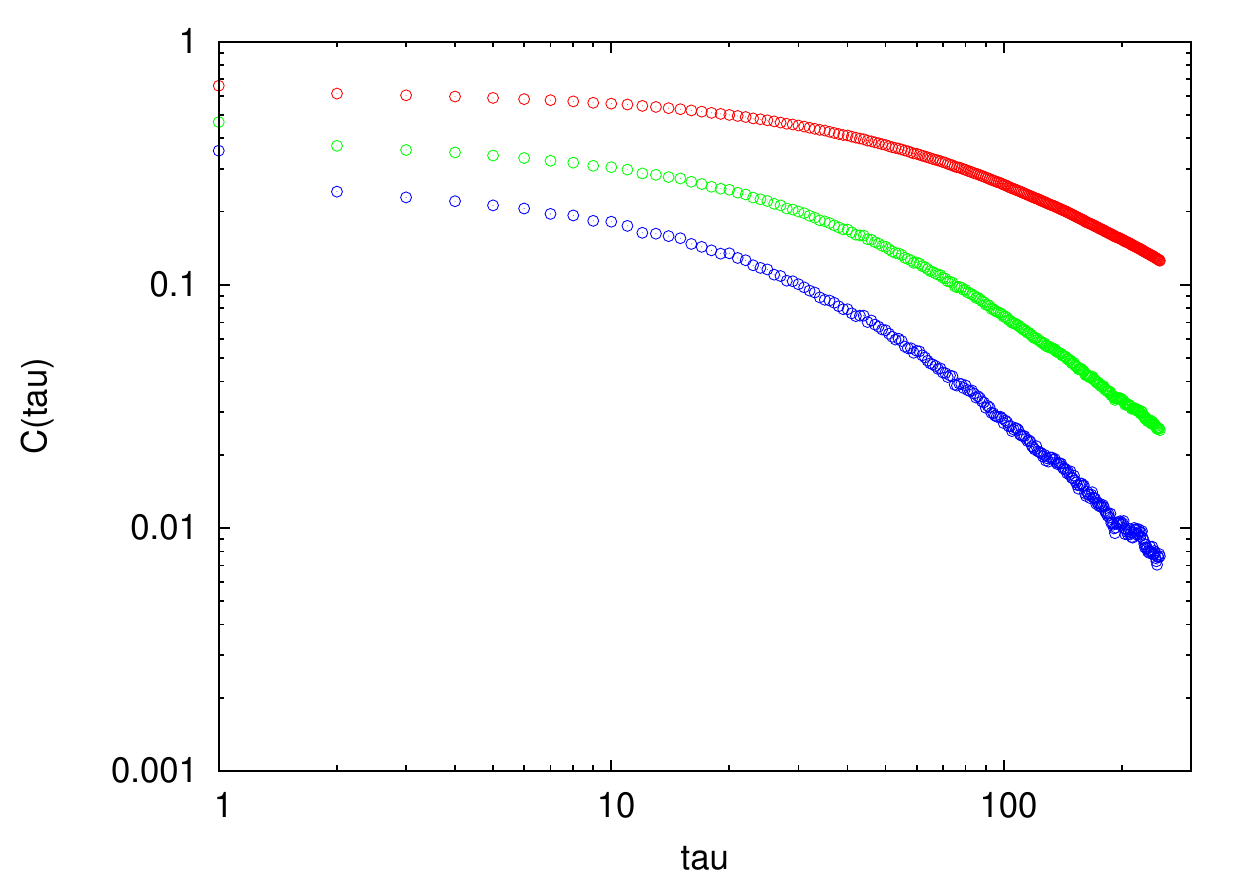}\par
	\parbox{70mm}{ \caption{\label{4545acor}\footnotesize $\langle C_\alpha(\tau) \rangle $ for $([0.45], (0.45))$} }
 \end{minipage} 
 \par  
 \begin{minipage}[t]{7 cm}
	\includegraphics[width= 70mm]{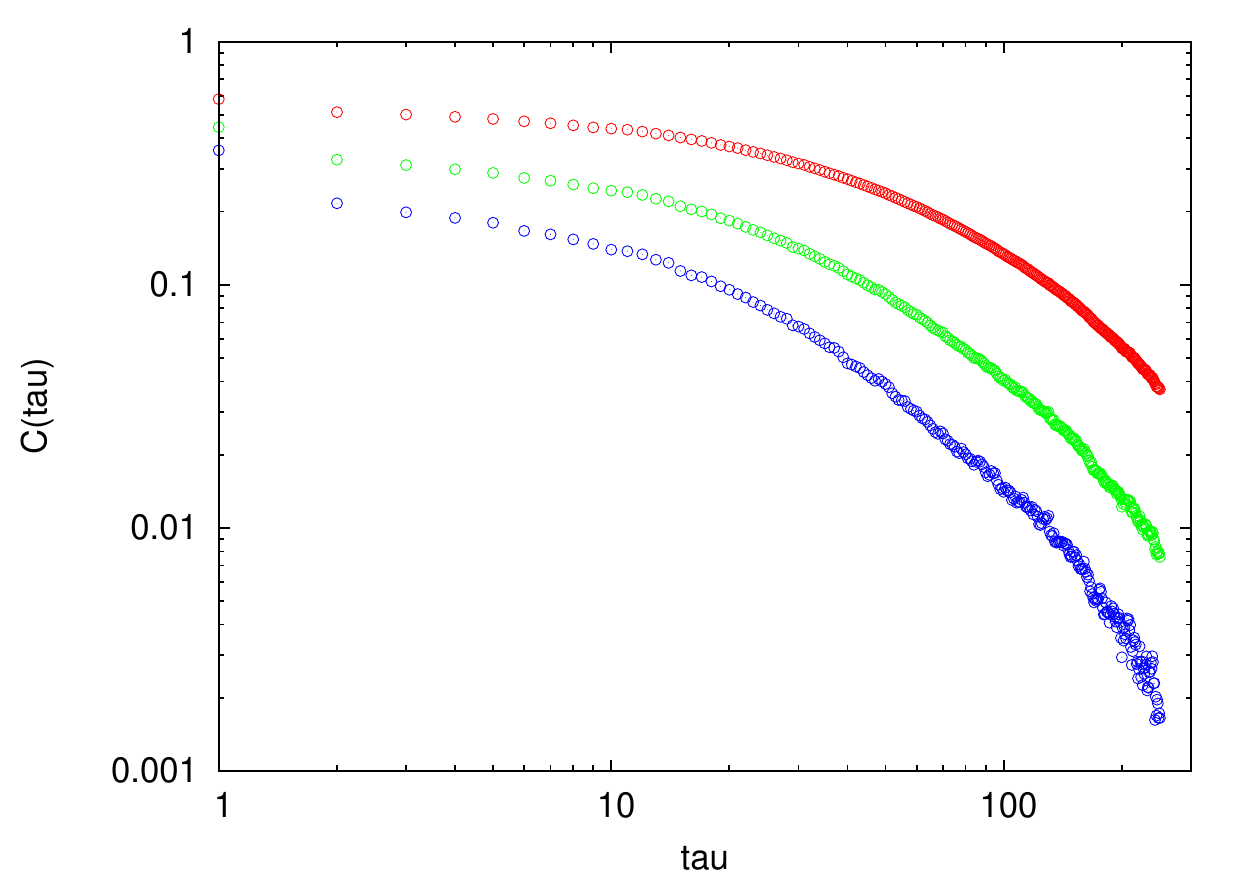}\par
	\parbox{50mm}{ \caption{\label{4065acor}\footnotesize $\langle C_\alpha(\tau) \rangle $ for $[0.45](0.65)$} }
 \end{minipage} \hskip 0.1cm
\begin{minipage}[t]{7 cm}
	 \includegraphics[width= 70mm]{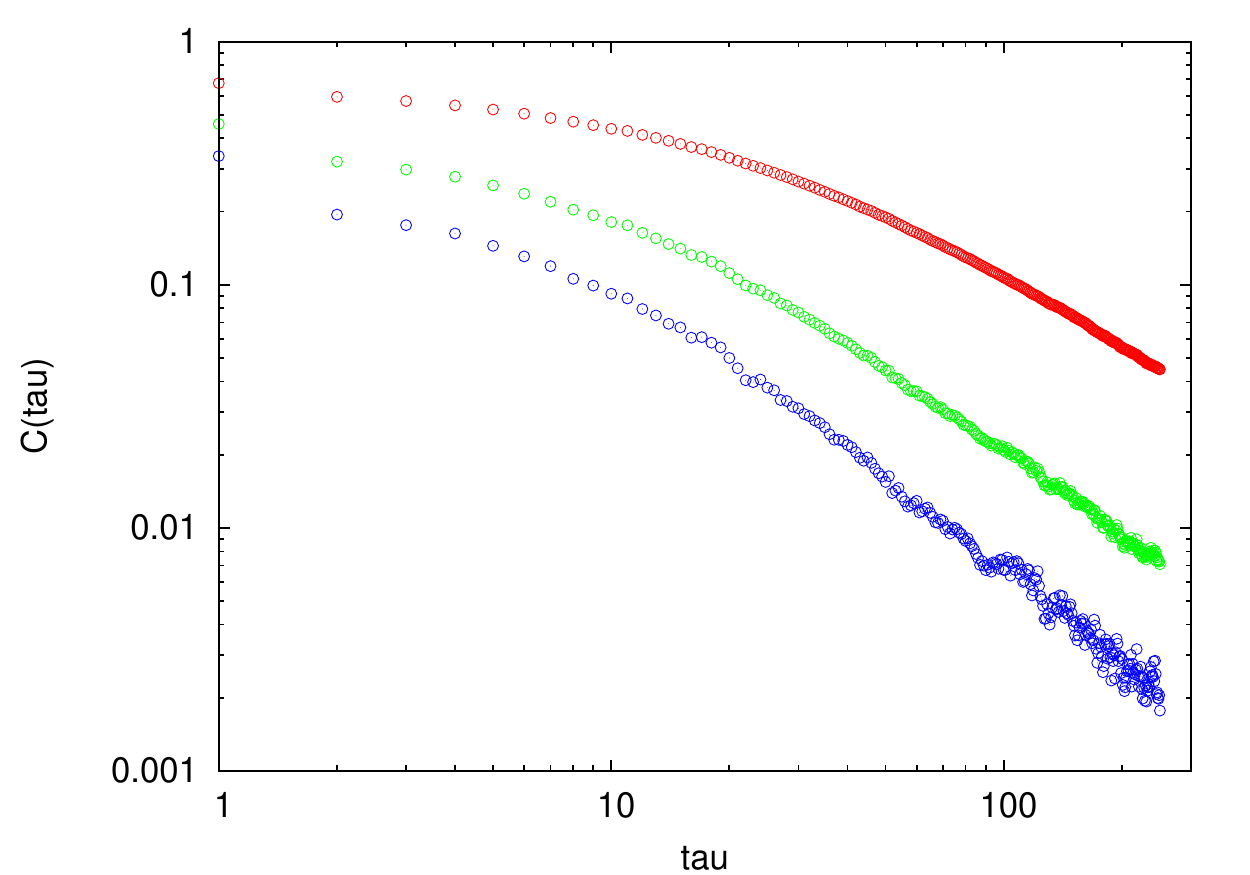}\par
	\parbox{50mm}{ \caption{\label{autocorrFluct}\footnotesize Fluctuations of $C_2(\tau)$ around the mean for the same parameters.} }
 \end{minipage} 
 \end{center}
\end{figure}

'Volatility clustering' is not an observable of return trails rather than depends on a particular model proposed. On the other hand autocorrelations of returns and their integer powers are observable. The slow decay of the autocorrelation of squared returns is often taken as a measure of 'volatility clustering'. In more general, empirical auto-correlation of integer powers of absolute returns $C_\alpha(\tau) = corr\bigg( |Z_{t+\tau}|^\alpha, |Z_t|^\alpha\bigg)$ have been found to decay slowly according to
$$
C_\alpha(\tau) \;\propto\; \tau^{-\gamma_\alpha} \qquad \tau \mbox{ large}, 
$$
where $\gamma_{1,2}$ was found to be in the range $[0.2,0.4]$. Ding and Granger remarked that typically this effect is largest for $\alpha = 1$, i.e. $\gamma_1 < \gamma_2$. Figures \ref{4525acor} to \ref{4065acor} show the autocorrelations $C_\alpha(\tau)$ for a market $([0.45], (b))$. For each parameter constellation $(a,b)$ we performed $1000$ runs. Respective $C_\alpha$ are averaged to give $\langle C_\alpha(\tau) \rangle$. $\langle C_1 (\tau) \rangle $ is red, $\langle C_2(\tau) \rangle $ is green, and $\langle C_3(\tau) \rangle $ is blue. 

Pictures show that $\gamma_1 \le \gamma_2 \le \gamma_3$, while differences between $\gamma_\alpha$ increase with increasing $b$. Figure \ref{autocorrFluct} shows the fluctuations of $\gamma_\alpha$ around the mean. The length of each bar represents the interval $[\langle C_\alpha(\tau) \rangle \pm 2 \: var(Z)]$. Fluctuations of $C_2(\tau)$ around the sample mean is seen to be moderate.

\subsection{Apparent multifractality}\label{multifract}

The singularity spectrum or the spectrum of local dimensions $D(\alpha)$ of a signal was introduced to give a characterization of the local smoothness structure of a time series in a statistical sense. Thereby $D(\alpha)$ can be regarded as the fractal dimension of the subset of points that possess the local scaling index $\alpha$. Investigations of pathwise regularities of empirical price trajectories revealed that the shape of the singularity spectrum is the same for a variety of assets. For review and an extended literature survey about multifractality in finance see \cite{LuxAusloos2001,AuslossIvanova2002}.
Models of a financial market as a 'true' multifractal system have been proposed by various authors quite recently, see \cite{MandelbrotFischerCalvet1997, LuxAusloos2001, Bacry2001a}. For a critical review see \cite{BorlandBouchaud2005}. \\

On the other hand, the existence of a non-trivial spectrum might not undoubtedly indicate that the system is truly multifractal itself. As shown in \cite{Berthelsen1994}, short time series of monofractal processes such as simple random walks may exhibit a nonlinear Hurst spectrum. 
This multiscaling behavior thus is related to finite sample sizes and to discretization rather than to the true multi fractal nature of the system under consideration. 
So-called {\em apparent multifractality} is also known from other systems, see \cite{BouchaudPottersMeyer2000}, including multiplicative random walk as well \cite{Redner1990}.
Moreover it is shown in \cite{BouchaudPottersMeyer2000} that multiscaling behavior can appear as a result of very long transient effects, induced by the long range nature of volatility correlations. In summary, multiscaling behavior might be apparent rather than a true system's signature. In \cite{Sornette1998} a regime of stochastic processes was shown to produce intermittency giving rise to fractal properties. However, the discussion about the possibly true multifractal nature of a financial market is still ongoing.
\begin{figure}[h] 
\setlength{\unitlength}{0.7cm}
\begin{center}
\begin{minipage}[t]{7 cm}
 \includegraphics[width=70mm]{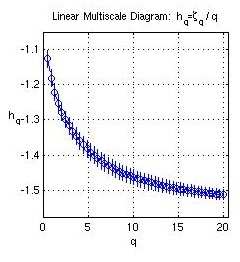}
\parbox{50mm}{ \caption{\label{MsSmodel}\footnotesize The Linear Multiscaling spectrum of the market $([0.4],(0.7))$, trail length 20.000.} }
 \end{minipage} \hskip 0.7cm  
 \begin{minipage}[t]{7 cm}
 \includegraphics[width= 76mm]{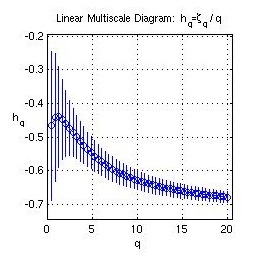}
\parbox{50mm}{ \caption{\label{MsSNIKKEI}\footnotesize The Linear Multiscaling spectrum of the NIKKEI, daily data from 1990 - 2005.} }
 \end{minipage} %\par %\hskip 0.5cm
 \end{center}
 \end{figure}
In the following we give some estimates for return trails simulated by our model. Its apparent multifractality might not come as a surprise anymore. Instead of $D$, the function $\zeta$ is shown which is essentially its Legendre transform, i.e. if $q\ge 0$ denote the order of some moment, it is known that 
$$
\zeta(q)-1 \; = \; {\cal L} \; D(\alpha) \; = \;\inf_{\alpha} \bigg( \alpha q - D(\alpha)\bigg). 
$$
For a monofractal, $\zeta(q) = \frac{q}{H}$, where $0 < H \le 2$, while deviations from linearity indicates the existence of multiple scale. The following Figures \ref{MsSmodel} and \ref{MsSNIKKEI} show the Linear Multiscaling Histogram of the scaling exponent $\zeta_q$, $h_q = \frac{\zeta_q}{q}$ in which linear behavior thus would be seen as a straight 
horizontal line\footnote{Algorithm by D. Veitch, P.Abry, P. Chainais in 2002}. The non-linear spectrum coincides with the existence of volatility clustering in our model.

\subsection{Cross over in the return distribution} \label{distribution}

The existence of multiple scales in the system implies that returns distributions are not invariant under the choice of different time-scales, i.e. one observes that distributions of returns with lags of the order of minutes, days, week, and
so forth deviate from each other, see \cite{SornetteCont1997, Gopikrishnan1999}. In fact distributions for long time scales like month' look quite Gaussian, while the distributions of High-Frequency data are convex shaped. Hence one observes a cross-over from concave to convex when considering different time scales. 

\begin{figure}[h]
\begin{center}
\includegraphics[width=13cm, height = 11cm]{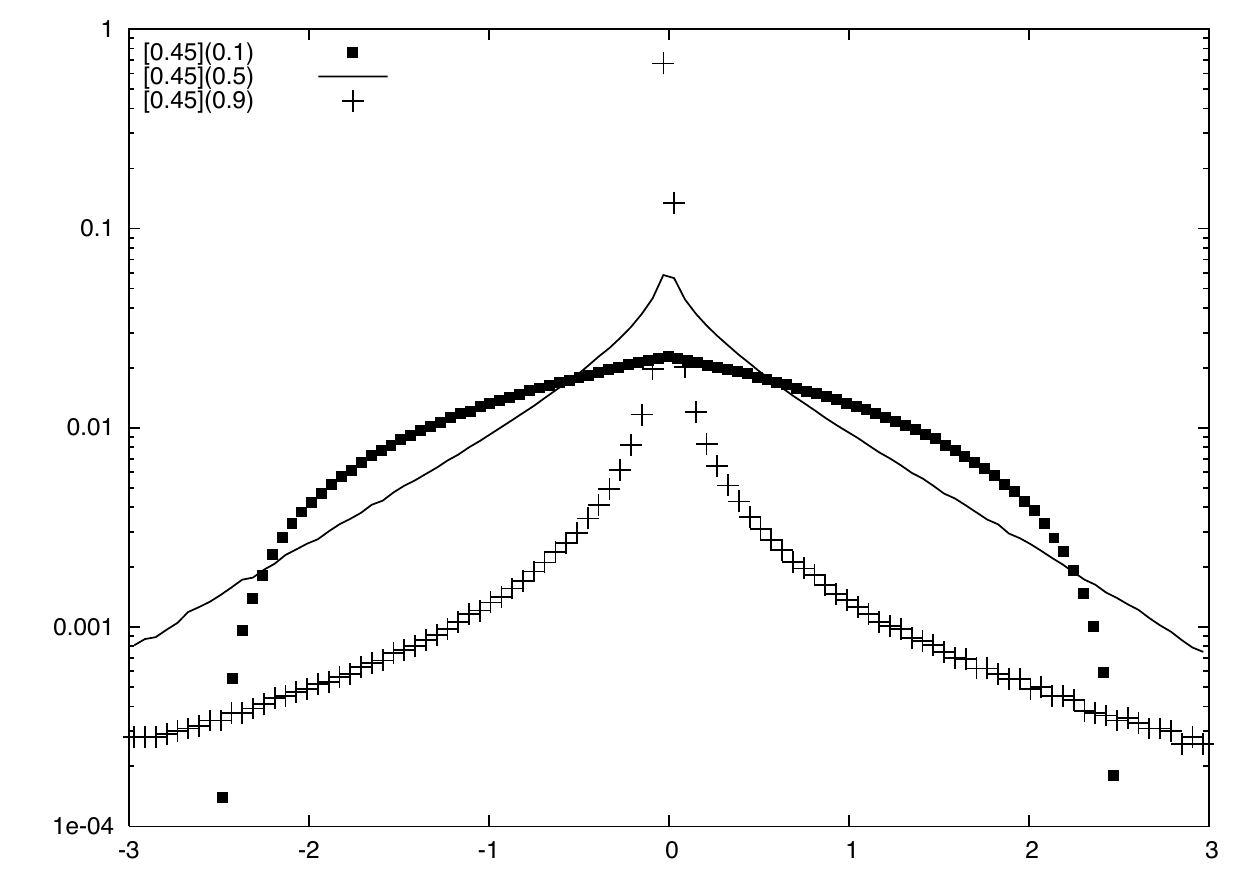}\par
\parbox{9.5cm}{\caption{ \label{crossover} \footnotesize Cross over on the market $([0.45],(\nu))$, $\nu = 0.1,0.5,0.9$ for 5.000 rum each of length 20.000.} }
\end{center}
\end{figure}

\begin{figure}[h]
\begin{center}
\includegraphics[width=13cm]{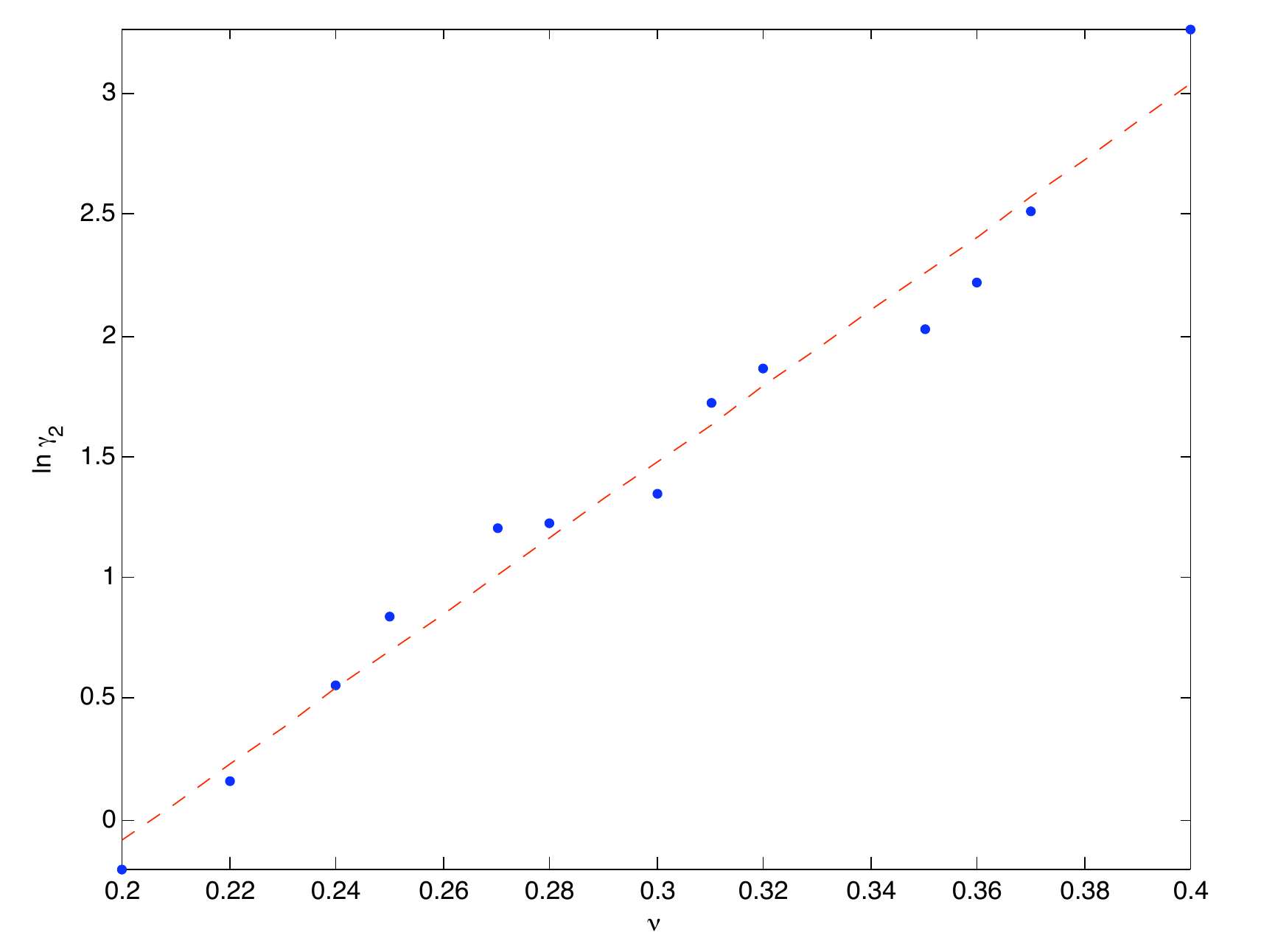}\par
\parbox{13cm}{
\caption{\label{excesskurtosis} \em \footnotesize Excess kurtosis $\gamma_2$ as a function of the fluctuation strength $b$ for simulated returns trails each of length 10.000 at fixed $r = 0.4$.}
}
\end{center}
\end{figure}

To better see extreme events and hence the tails of the distribution, simulations in Fig \ref{crossover} were done for longer runs with trail length 20.000. The pdf in Figure \ref{crossover} is the averaged over 1.000 runs. As seen in Fig \ref{crossover} the distributions of simulated return trails show a cross over from a concave Gaussian-like distribution for small $b$ to a convex distribution for large $b$, while for some $b$ the distribution is a Laplacian.  This cross over typically appears when further separating the two time scales from each other. Figure \ref{excesskurtosis} suggests that in this model there is a simple relation between the strength $b$ of fluctuations and the excess kurtosis $\gamma_2$
$$
\gamma_2 \propto e^b, \qquad b > 0
$$

\section{Conclusion}
\begin{figure}[h]
\begin{center}
\includegraphics[width=15cm]{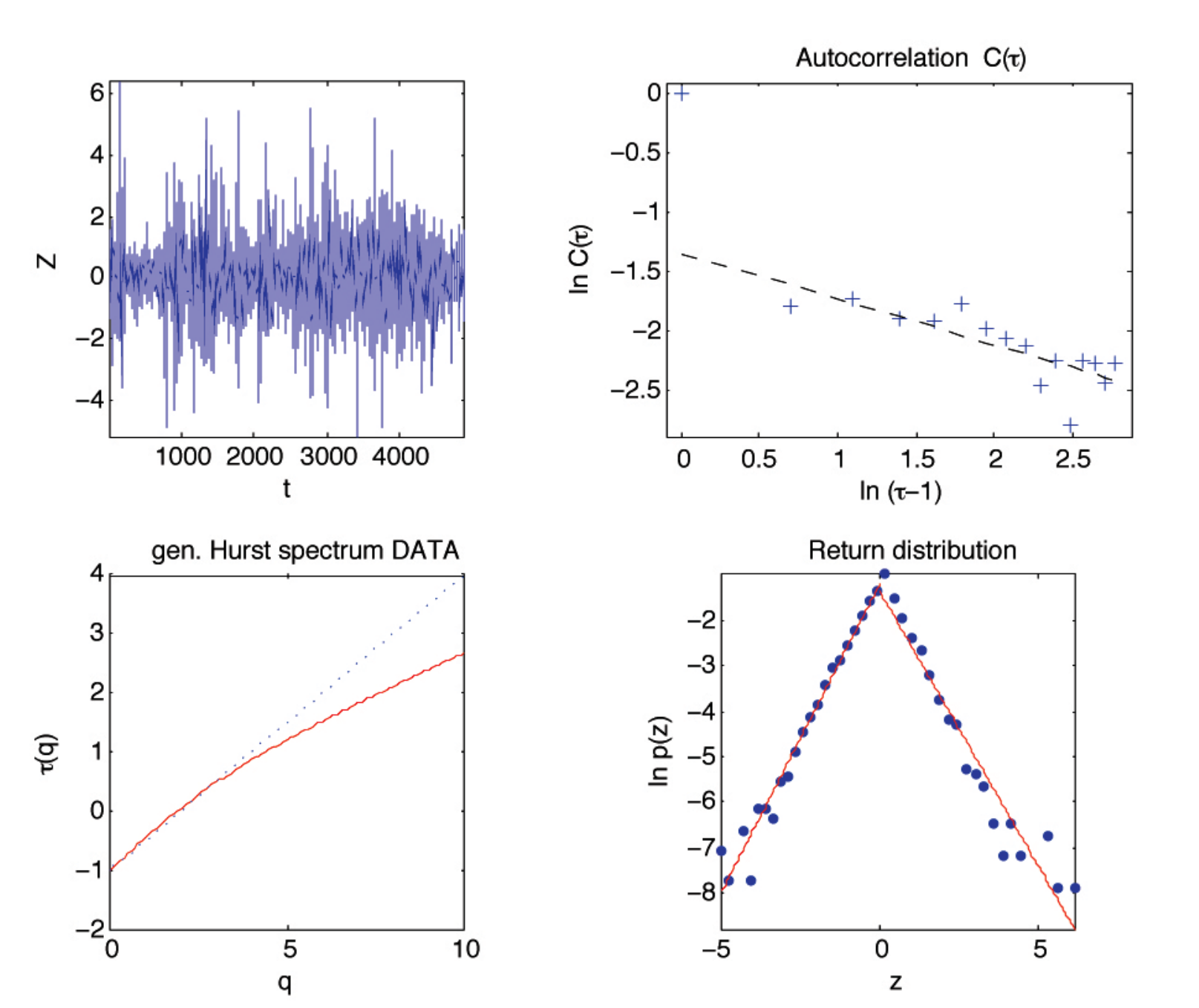}
\parbox{15cm}{
\caption{\label{NIKKEI} \em \footnotesize Summary of stochastic properties of the NIKKEI 250, daily returns from 10/30/1986 to 12/30/2005}
}
\end{center}
\end{figure}

\begin{figure}[h]
\begin{center}
\includegraphics[width=15cm]{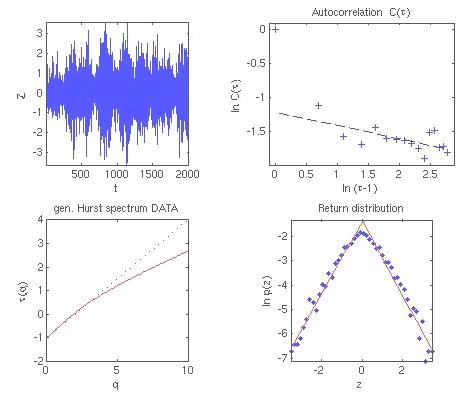}
\parbox{12cm}{

\caption{\label{ALL} \em \footnotesize Summary of stochastic properties in the market $([0.4],(0.44))$}
}
\end{center}
\end{figure}

Figures \ref{NIKKEI} and \ref{ALL} summarize stochastic properties of time series from the NIKKEI 250, daily data, and those generated by our model. Properties include volatility clustering, seen as the slow decay of auto correlations in squared returns, the heavy tailed distribution of returns, as well as the non-linear spectrum of singularities in the return trail. These finding serve as major stylized fact in empirical asset returns. Given these results of the model it might be interesting to resume what this model represents. 

Our model does not mimic the variety of individual investors and their trading activities on the market. It is a model for the dynamics of prices. While the assumption that all investors are fully invested all the time is unrealistic, prices derived under this hypothesis serve as a zero-order approximation provided that the degree for cash investment is sufficiently small. Basic result remain qualitatively unaffected by this approximation.\\

Since prices are macro observables, this is a model for dynamics {\em on the aggregate level}. Since macroscopic typical properties are not necessarily up-stream {\em analogies} of microscopic properties, there is no {\em a priori} reason to assume that dynamics on the macro level should be similar to that on the microscopic level. Instead the model considers the financial market as a complex system whose entire dynamics can be described by the coupled dynamics of two dynamical regimes which differ by their time scales, a slow one and a fast one. Accordingly our model has two dynamical components, a static one and a faster, noisy one. This is quite similar to what is done in a mean field approximation, in lowest order. To even further simply the model we assumed that the dynamics of the fast component is linear and random. 
The fast component fluctuates with an amplitude $|b|$. The amplitude of these fluctuations significantly affects statistical properties of the entire system. Particularly, if the noisy component is almost constant, the system behaves quite regularly in the sense that returns are close to white noise. This situation changes drastically if the noisy component becomes stronger. In this case, volatility clustering becomes more pronounced, i.e. the decay of auto correlations of integer powers of absolute returns slows down. Correspondingly the multifractal signature of return time series becomes more significant. In parallel, one observes a cross over in the return distribution from a concave, Gaussian-like shape over a Laplacian to a convex shape when the strength of fluctuations in the system increases.\\

These findings are not restricted to only a particular set of parameter values but emerge and unfold over the entire range of the model, i.e. $0 < |b| \le 1$. This suggests that these properties have their roots in the general structure of the model. Recall that our model is a multiplicative random process whose stochastic growth rate depends on the current price and thereby establishes a negative, multiplicative feedback. This structure comes quite naturally from simple economic considerations. It can be shown that even a much simpler model having the same feedback structure as the presented one creates the qualitatively the same stylized facts \cite{Reimann2007b}. Hence it seems to be this particular structure, which is responsible for the properties of the system. The claim is that {\em any} model exhibiting this feedback structure creates statistical properties qualitatively similar to empirical stylized facts, including volatility clustering, apparent multi fractality, and fat-tailed distributions. \\

The core might be the following: Time in general and the coupling of processes in particular appear to be essential for understanding the behavior of a financial market. Hence mentioning processes and their temporal coupling should not be ignored in model building of economic systems.


\begin{thebibliography}{10}

\bibitem{Amir2005}
R.~Amir, I.~V. Evstigneev, T.~Hens, and K.~R. Schenk-Hopp\'e.
\newblock Market selection and survival of investment strategies.
\newblock {\em J. Math. Economics}, 41:105--122, 2005.

\bibitem{AuslossIvanova2002}
M~Ausloos and K.~Ivanova.
\newblock Multifractal nature of stock exchange prices.
\newblock {\em Computer Physics Communications}, 147:582 -- 585, 2002.

\bibitem{Barndorff-Nielsen1978}
O.~E. Barndorff-Nielsen.
\newblock Hyperbolic distributions and distributions on hyperbolae.
\newblock {\em Scand. J. Statist.}, 5:151--157, 1978.

\bibitem{BibbySorensen1997}
B.~M. Bibby and M.~Sorensen.
\newblock A hyperbolic diffusion model for stock prices.
\newblock {\em Finance and Stochastics}, 1:25 -- 42, 1997.

\bibitem{BlumeEasley1992}
Lawrence Blume and David Easley.
\newblock Evolution and market behavior.
\newblock {\em Journal of Economic Theory}, 58(1):9--40, 1992.

\bibitem{BouchaudPottersMeyer2000}
J.-P. Bouchaud, M.~Potters, and M.~Meyer.
\newblock Apparent multifractality in financial time series.
\newblock {\em Phys. J. B}, 13(3):595--599, 2000.

\bibitem{BouchaudPotters2003}
Jean-Philippe Bouchaud and Marc Potters.
\newblock {\em Theory of Financial Risk and Derivative Pricing}.
\newblock Cambridge University Press, 2003.

\bibitem{Cont2001}
R.~Cont.
\newblock Empirical properties of asset returns: stylized facts and statistical
  issues.
\newblock {\em Quantitative Finance}, 1:223--236, 2001.

\bibitem{Bacry2001a}
J.F.~Muzy E.~Bacry, J.~Delour.
\newblock Multi-fractal nature of stock exchange prices.
\newblock {\em Phys. Rev. E}, 64, 2001.

\bibitem{Berthelsen1994}
A.~Berthelsen et~al.
\newblock Effective multifractal sprectrum of a random walk.
\newblock {\em Phys. Rev. E}, 49:1860, 1994.

\bibitem{BorlandBouchaud2005}
L.~Borland et~al.
\newblock The dynamics of financial markets --- mandelbot's multifractal
  cascades, and beyond.
\newblock Technical report, Arxiv preprint cond-mat/0501292, 2005.

\bibitem{LuxAusloos2001}
Th. Lux and M.~Ausloos.
\newblock Market fluctuations i : Scaling, multi-scaling and their possible
  origins.
\newblock In {\em The Science of Disasters: Scaling Laws Governing Weather,
  Body, Stock-Market Dynamics}, pages 377--413. Springer, 2001.

\bibitem{Mandelbrot1962}
B.~Mandelbrot.
\newblock Paretian distributions and income maximization.
\newblock {\em Quarterly Journal of Economics}, 76:57--85, 1962.

\bibitem{MandelbrotFischerCalvet1997}
B.~Mandelbrot, Fischer A., and Calvet L.
\newblock Multifractal model of asset returns.
\newblock Technical Report Discussion Paper \# 1164, Cowles Foundation, 1997.

\bibitem{MantegnaStanley2000}
Rosario~N. Mantegna and H.~Eugene Stanley.
\newblock {\em An Introduction to Econophysics: Correlations and Complexity in
  Finance}.
\newblock Cambridge University Press, 2000.

\bibitem{Gopikrishnan1999}
Gopikrishnan P., Plerou V., Amaral L.A.N., Meyer M., and Stanley H.E.
\newblock Scaling of the distribution of fluctuations of financial market
  indices.
\newblock {\em Phys. Rev. E}, 60:5305--5316, 1999.

\bibitem{Prause1998}
K.~Prause.
\newblock {\em The generalized hypervolic model}.
\newblock PhD thesis, University of Freiburg, 1998.

\bibitem{MantegnaStanley1994}
E.~Stanley R.~N.~Mantegna.
\newblock Stochastic process with ultraslow convergence to a gaussian: The
  truncated l\`evy flight.
\newblock {\em Phys. Rev. Lett.}, 73:2946, 1994.

\bibitem{Redner1990}
S.~Redner.
\newblock Multiplicative random walks: an elementary tutorial.
\newblock {\em Am. J. Phys.}, 58:267, 1990.

\bibitem{Reimann2007b}
S.~Reimann.
\newblock Price dynamics from a simple multiplicative random process model:
  stylized facts and beyond?
\newblock {\em European Physics Journal B}, B(56):381 -- 394, 2007.

\bibitem{Reimann2007a}
S.~Reimann and A.~Tupak.
\newblock Prices are macro observables: Stylized facts from evolutionary
  finance.
\newblock {\em Computational Economics}, 29(3-4):313 -- 331, 2007.

\bibitem{Sorensen2003}
M.~Sorensen.
\newblock Hyperbolic processes in finance.
\newblock In S.~Rachev, editor, {\em Handbook of heavy Tailed Distributions in
  Finance}, pages 211--248. Elsevier Science, 2003.

\bibitem{Sornette1998}
D.~Sornette.
\newblock Linear stochastic dynamics with nonlinear fractal properties.
\newblock {\em Physica A}, 250:295 -- 314, 1998.

\bibitem{SornetteCont1997}
D.~Sornette and R.~Cont.
\newblock Convergent multiplicative processes repelled from zero: Power laws
  and truncated power laws.
\newblock {\em J. Phys. I France}, 7:431--444, 1997.

\end{thebibliography}
\end{document}